\documentclass[journal=mamobx,manuscript=article,layout=onecolumn]{achemso}
\setkeys{acs}{articletitle = true}
\AbstractOn
\SectionNumbersOn

\usepackage[version=3]{mhchem} % Formula subscripts using \ce{}
\usepackage{xcolor}
\usepackage{amsmath}
\usepackage{amssymb}
\usepackage{natmove}
\author{Francesco\@ Puosi}
\email{francesco.puosi@pi.infn.it}
\affiliation{Dipartimento di Fisica ``Enrico Fermi'', 
	Universit\`a di Pisa, Largo B.\@Pontecorvo 3, I-56127 Pisa, Italy}
	\alsoaffiliation{INFN, Sezione di Pisa, Largo B. Pontecorvo 3, Pisa I-56127, Italy}
	\altaffiliation{These authors contributed equally to this work.}

\author{Antonio\@ Tripodo}
\affiliation{Dipartimento di Fisica ``Enrico Fermi'', 
	Universit\`a di Pisa, Largo B.\@Pontecorvo 3, I-56127 Pisa, Italy}
	\altaffiliation{These authors contributed equally to this work.}

\author{Marco\@ Malvaldi}
\affiliation{Dipartimento di Fisica ``Enrico Fermi'', 
	Universit\`a di Pisa, Largo B.\@Pontecorvo 3, I-56127 Pisa, Italy}

\author{Dino\@ Leporini}
\email{dino.leporini@unipi.it}
\affiliation{Dipartimento di Fisica ``Enrico Fermi'', 
	Universit\`a di Pisa, Largo B.\@Pontecorvo 3, I-56127 Pisa, Italy}
\alsoaffiliation{Istituto per i Processi Chimico-Fisici-Consiglio Nazionale delle Ricerche (IPCF-CNR), via G. Moruzzi 1, I-56124 Pisa, Italy}

\title[An \textsf{achemso} demo]
{Johari-Goldstein heterogeneous dynamics in a model polymer}

\abbreviations{IR,NMR,UV}

\begin{document}

%%%%%%%%%%%%%%%%%%%%%%%%%%%%%%%%%%%%%%%%%%%%%%%%%%%%%%%%%%%%%%%%%%%%%
%% The "tocentry" environment can be used to create an entry for the
%% graphical table of contents. It is given here as some journals
%% require that it is printed as part of the abstract page. It will
%% be automatically moved as appropriate.
%%%%%%%%%%%%%%%%%%%%%%%%%%%%%%%%%%%%%%%%%%%%%%%%%%%%%%%%%%%%%%%%%%%%%
\begin{tocentry}
\includegraphics{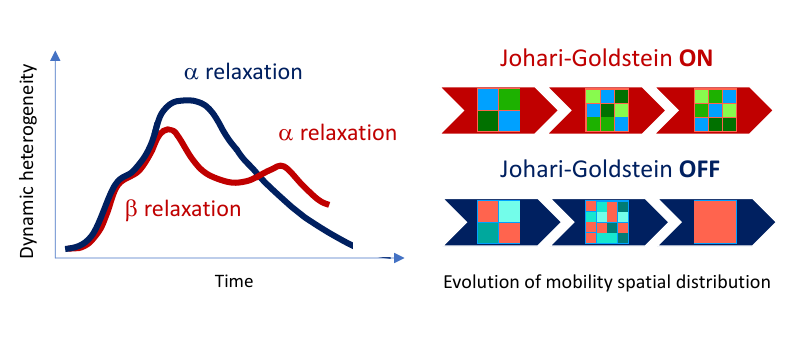}
\end{tocentry}

%%%%%%%%%%%%%%%%%%%%%%%%%%%%%%%%%%%%%%%%%%%%%%%%%%%%%%%%%%%%%%%%%%%%%
%% The abstract environment will automatically gobble the contents
%% if an abstract is not used by the target journal.
%%%%%%%%%%%%%%%%%%%%%%%%%%%%%%%%%%%%%%%%%%%%%%%%%%%%%%%%%%%%%%%%%%%%%
\begin{abstract}
 The heterogeneous character of the Johari-Goldstein (JG) relaxation is evidenced by molecular-dynamics simulation of a model polymer system. A double-peaked evolution of dynamic heterogeneity (DH), with maxima located at JG and structural relaxation time scales, is observed and mechanistically explained. {The short-time single-particle displacement during JG relaxation weakly correlates with the long-time one observed during structural relaxation}.
 \end{abstract}

%%%%%%%%%%%%%%%%%%%%%%%%%%%%%%%%%%%%%%%%%%%%%%%%%%%%%%%%%%%%%%%%%%%%%
%% Start the main part of the manuscript here.
%%%%%%%%%%%%%%%%%%%%%%%%%%%%%%%%%%%%%%%%%%%%%%%%%%%%%%%%%%%%%%%%%%%%%
\section{Introduction}

Avoiding crystallization, polymers and liquids freeze into a microscopically disordered solid-like state, a glass  \cite{DebenedettiBook}. 
On approaching the glass transition, molecular rearrangements occur via both the primary mode, referred to as structural or $\alpha$ relaxation, and the faster secondary ($\beta$) processes as evidenced by mechanical, electrical, and thermal properties of materials  \cite{WilliamsMcCrum, AngelNgai00,NgaiBook}.
{ There is wide interest in the $\beta$ relaxation, as attested by the} large number of experiments, as well as phenomenological and theoretical studies, and simulations
\cite{JOHARI70,Ngai98,NgaiPaluchClassificationSecondaryJCP04,NgaiBook,Capaccioli12,CiceroneDynHeterogJohariGoldsteinPRL14,yu2017}. 

Here, we focus on linear polymers where there are no side groups and the secondary relaxations are  usually ascribed to some movement of short lengths of the main chain, like to limited vibrational oscillation about their mean position or crankshaft motion \cite{WilliamsMcCrum,Boyd:1974lq,Paul:1997wd,Struik98}. The secondary relaxation in linear polymers is thought to be a genuine manifestation of the Johari Goldstein (JG) $\beta$-relaxation \cite{JOHARI70}, 
{ a special class of secondary relaxations  having strong connections  to the $\alpha$-relaxation in properties, and 
advocated to be a universal feature of the glass transition
 \cite{Ngai98,NgaiPaluchClassificationSecondaryJCP04, NgaiBook,Capaccioli12,GOLDSTEIN_JNCS11}}.
 
 It was early noted that  the molecular rearrangements giving rise to JG relaxation process are similar to those that are responsible for the glass transition itself  \cite{Johari76}. In particular, correlation between JG and primary relaxation were reported in both molecular liquids and polymers \cite{BERSHTEIN199441,NgaiAlfaBetaCor98,Struik98,BohmerAlfaBetaPRL06_Cit85}. It was argued that the JG relaxation, like the alpha, involve transitions between metabasins in the energy landscape \cite{GoldsteinBetaDynbamicalHetJCP10,CiceroneMetabasinJG_JCP17} {, whereas other approaches suggest that JG processes can be
interpreted in structural glasses as transitions between sub-basins belonging to a same metabasin  \cite{CharbonneauKurchanParisiZamponoNatComm14}.}
 
  Furthermore, a number of experimental results indicate that the JG relaxation is sensitive to the thermodynamic variables underlying the glass transition \cite{NgaiPaluchClassificationSecondaryJCP04, NgaiBook}, mimicking the $\alpha$ relaxation, being strongly pressure dependent and showing the invariance of the ratio $\tau_\alpha$/$\tau_\beta$ to variations of pressure and temperature, keeping $\tau_\alpha$ constant {\cite{Capaccioli12}}. This led to the conclusion that the JG relaxation is precursor to structural relaxation and viscous flow, having a slower dynamics due to cooperativity involving many body dynamics  { \cite{Ngai98,Capaccioli12,Johari:2019dw,CapaccioliNgaiJGPhilMag20}. This has been recently substantiated by studying the invariance of the relation between $\alpha$  relaxation and $\beta$ relaxation in metallic glasses to variations of pressure and temperature by molecular dynamics simulations combined with the dynamic mechanical spectroscopy method \cite{NgaiMetallicGlassPRB20}.}
 Further support to the view that primary and JG relaxation are closely related also come from the evidence that the JG relaxation shows both a very broad distribution of relaxation times \cite{AngelNgai00,Johari:2019dw},{ \cite{BedrovSmithJPolymSci07}} and cooperative dynamics, indicated by { simulations in non-polymeric liquids \cite{KarmakarSastryCooperativeBetaPRL16}, experiments \cite{Capaccioli12,CapaccioliNgaiJGPhilMag20}}. 
  
  {
The existence of spatial correlations between dynamic fluctuations - in short, dynamic heterogeneity (DH) - has been revealed by experiments  and numerical studies, e.g., see the reviews in~\cite{SILLESCURevDynHet99,Ediger00,Richert02,BerthieRev,SastryLengthScalesRepProgrPhys15,Napolitano:2013rc,Napolitano_2017}. In particular, the presence of DH in the $\alpha$ relaxation regime has been studied in bulk polymers, by e.g. multidimensional NMR  \cite{TrachtSpiessDynHetPRL98} and simulations \cite{Colmenero_MD_DH_PRE02}, { and tuned by nanoconfinement \cite{Napolitano:2013rc,Napolitano_2017}}. 
Of particular interest to the present study are the findings that DHs at both $\alpha$ and $\beta$ relaxation time scales have been reported in colloids as clusters of faster-moving particles \cite{WeeksDHColloidScience00} and numerical studies found heterogeneous dynamics of JG relaxation \cite{FragiadakisRolandAlfaBetaPRE17}, supporting previous suggestions \cite{GoldsteinBetaDynbamicalHetJCP10}. On even shorter time scales, picosecond DH - observed by incoherent quasielastic neutron scattering - allowed the evaluation of the characteristic time scale of primary relaxation of molecular liquids \cite{CiceroneDynHeterogJohariGoldsteinPRL14}. Furthermore, it has been noted that JG involves a broad distribution of processes with those occurring at longer times being characterized by a longer length scale \cite{Capaccioli12,CapaccioliNgaiJGPhilMag20}.}
 
A familiar tool to expose and quantify DH is the non-gaussian parameter (NGP) \cite{HansenMcDonaldIIIEd}:
\begin{equation}
\alpha_2(t) = \frac{3}{5} \frac{\langle r^4(t) \rangle}{ \langle r^2(t) \rangle^2} - 1
\end{equation}
where $r(t)$ and $\langle \cdots \rangle$ denote the modulus of the particle displacement in a time $t$ and the ensemble average, respectively. NGP vanishes if the displacement  is accounted for by a {\it single}, i.e. spatially homogeneous, gaussian random process \cite{HansenMcDonaldIIIEd}. Instead, if dispersion is present and  the individual particles undergo  {\it distinct} motions (even if with gaussian features), NGP is positive, $\alpha_2(t) >0$ \cite{ZornNonGaussianPolybutadienePRB97}.  Notably, NGP is accessible to experiments e.g. by confocal microscopy  \cite{WeeksDHColloidScience00} and neutron scattering \cite{ZornNonGaussianPolybutadienePRB97}. { Recently, JG $\beta$ relaxation has been  resolved by studying the isochronal NGP in simulations of metallic glasses \cite{NgaiMetallicGlassPRB20}}.

 It must be pointed out that a distribution of relaxation times, as observed in JG relaxation \cite{AngelNgai00,Johari:2019dw}, does not imply necessarily non-gaussian displacements { - i.e. non vanishing NGP, the customary criterion to identify DH -}, e.g., see the Rouse model which predicts {\it multiple} relaxation times of an unentangled single chain with {\it gaussian} particle displacements \cite{DoiEdwards}.

In this work we report results from extensive molecular-dynamics (MD) simulations of a polymer model  melt proving, when JG relaxation is present, a bimodal double peaked NGP leading to two distinct DH growths in the $\beta$ and $\alpha$ relaxation regimes. A mechanistic microscopic explanation is provided.   Our findings capture a new  correspondence between  JG and  structural relaxation in polymers.

\section{Model and Numerical Methods}
\label{methods}

\begin{figure}
	\centering
	\includegraphics[width=0.40\linewidth]{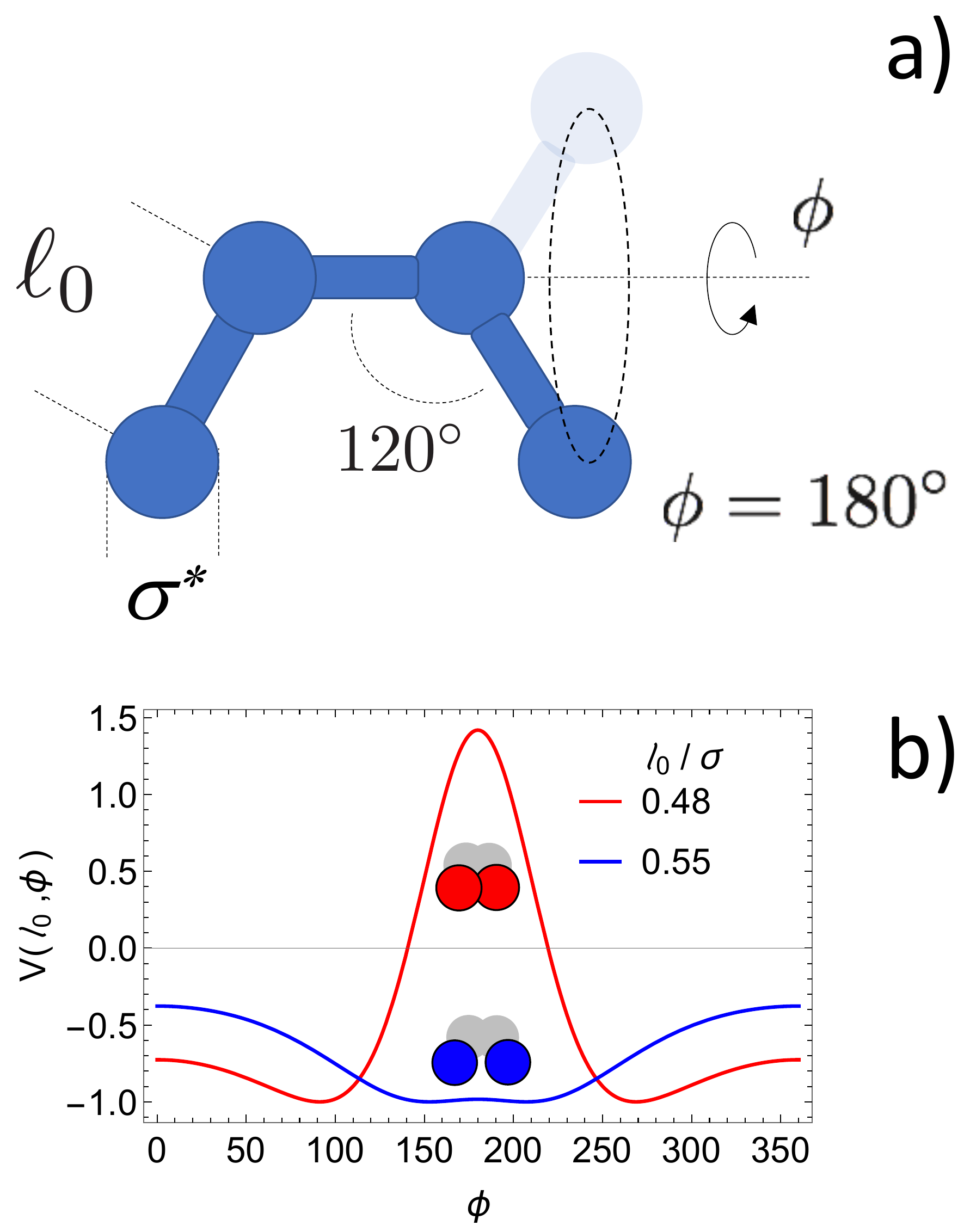}
	\caption{(a): fragment of the linear chain. The bond length is $ \simeq \ell_0$ and the adjacent bonds form an angle $\simeq 120^\circ$. The monomer size is about $\sigma^\star = 2^{1/6} \, \sigma$ where $\sigma$ is the length scale of the LJ potential (see SI for details). The dihedral angle $\phi$ is indicated. (b): illustration of the torsional barrier in the interacting potential $V(\ell_0, \phi)$ between the first and the last monomer of the fragment due to the LJ repulsion in {\it cis} configuration when $\ell_0 < 0.5 \, \sigma$.}
	\label{bond}
\end{figure}

We adopt a variant of coarse-grained models of linear polymers having nearly fixed bond length and bond angles constrained to $120^{\circ}$, Fig.\ref{bond}a \cite{BedrovPRE2005,BedrovJNCS2011,FragiadakisRolandMM17}. Details are given as Supporting Information (SI).  All the data presented in this work are expressed in reduced MD units, in particular lengths are in units of the length scale $\sigma$ of the Lennard-Jones (LJ) potential.  An interesting aspect of our variant is that, even if the force field does not include a torsional potential, { thus saving computing times,} an effective torsional barrier occurs for  $l_0 < 0.5 \sigma$, see Fig. \ref{bond}b, due to the LJ repulsion between the farther two monomers in a chain fragment of four monomers, see Fig. \ref{bond}a. We focus on $l_0=0.48 \,\sigma$ and $l_0=0.55 \,\sigma$ leading to considerable or missing torsional barrier, respectively, see Fig. \ref{bond}b. { It is worth noting that the assessment of the interplay of the intra-chain torsional barriers and other barriers, e.g. the ones arising from packing constraints, is a difficult task and goes beyond the purposes of the present work.}
	
\section{Results and discussion}
\label{res_disc}

\subsection{Detection of JG process by bond reorientation}
\label{JGBond}

Experiments and simulations demonstrated that orientational correlation functions are sensitive to detect and resolve secondary motions  \cite{Richter_PhysB1997,tolle,Fragiadakis_PRE2014,FragiadakisRolandAlfaBetaPRE17,TripodoBetaPolymers2020}, in particular, the reorientation of the chain bonds \cite{FragiadakisRolandAlfaBetaPRE17,TripodoBetaPolymers2020}. Let us define the bond correlation function (BCF) $C(t)$ as \cite{CapacciEtAl04}:
\begin{equation}
C(t) = \langle \cos \theta(t) \rangle
\label{defct}
\end{equation}
where $\theta(t)$ is the angle spanned in a time $t$ by the unit vector along a generic bond of a chain. An average over all the bonds is understood.  Starting from the unit value, BCF decreases in time, finally vanishing at long times when the bond orientation has spanned all the unit sphere.

\begin{figure}
	\centering
	\includegraphics[width=0.40\linewidth]{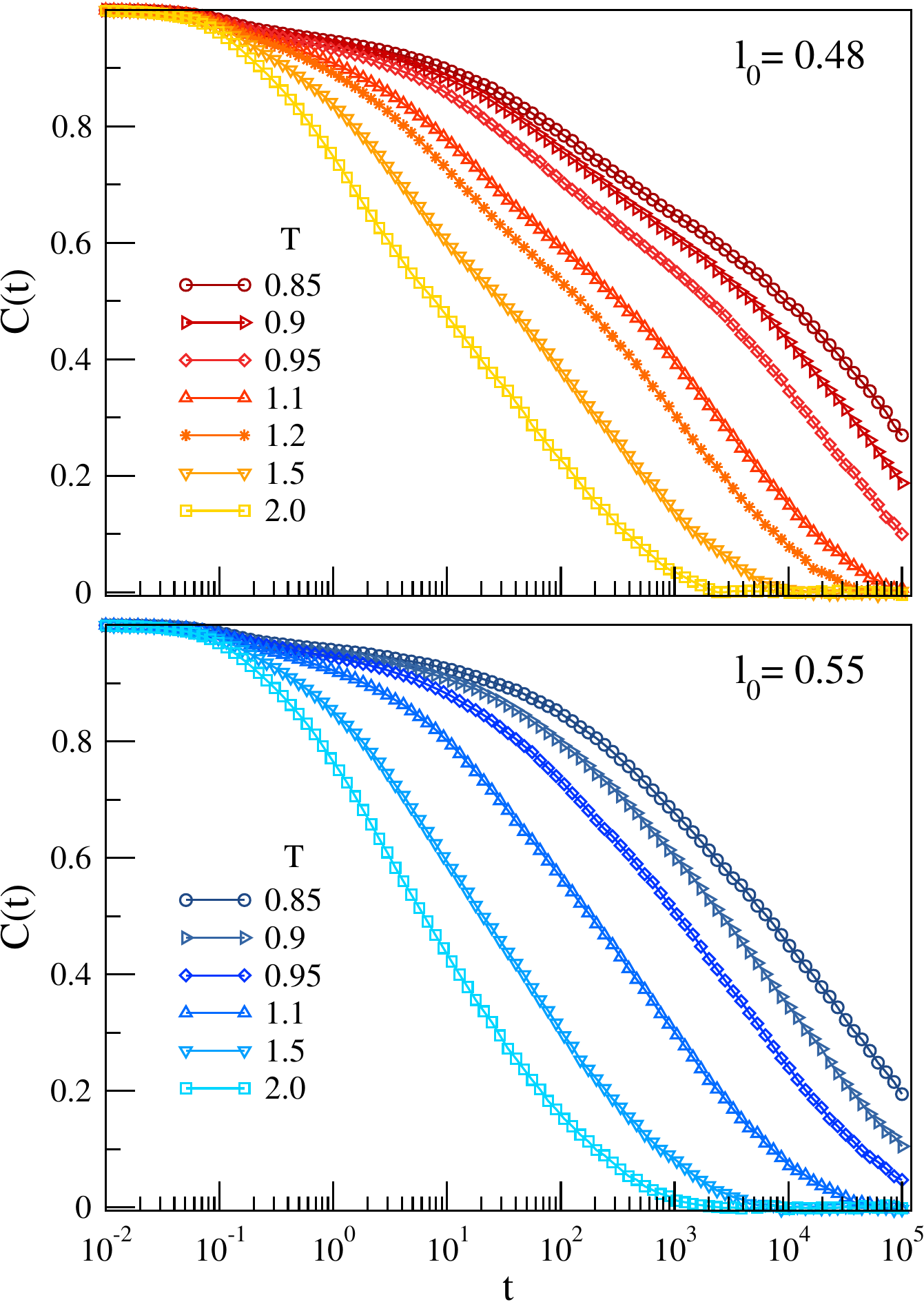}
	\caption{BCF with  bond length $\ell_0 = 0.48$ (top) and $\ell_0 = 0.55$ (bottom)  at different temperatures. If $l_0=0.48$, a clear two-step decay --- evidencing two distinct relaxations--- is observed.}
	\label{ct}
\end{figure}

Fig. \ref{ct}  plots BCF for the bonds with length $l_0=0.48 \,\sigma$ and $l_0=0.55 \,\sigma$. In agreement with previous MD studies on similar model polymers \cite{BedrovJNCS2011}, it is seen that in the presence of shorter bonds BCF exhibits a characteristic two-step decay (in addition to the initial decay  for $t \lesssim 1$), signaling the presence of two distinct  relaxation processes interpreted as the JG and structural relaxation \cite{BedrovPRE2005,BedrovJNCS2011,TripodoBetaPolymers2020}. { The relaxation maps of the JG and the $\alpha$ processes for the present model were reported elsewhere \cite{TripodoBetaPolymers2020}}.

\begin{figure}
	\centering
	\includegraphics[width= 0.40\linewidth]{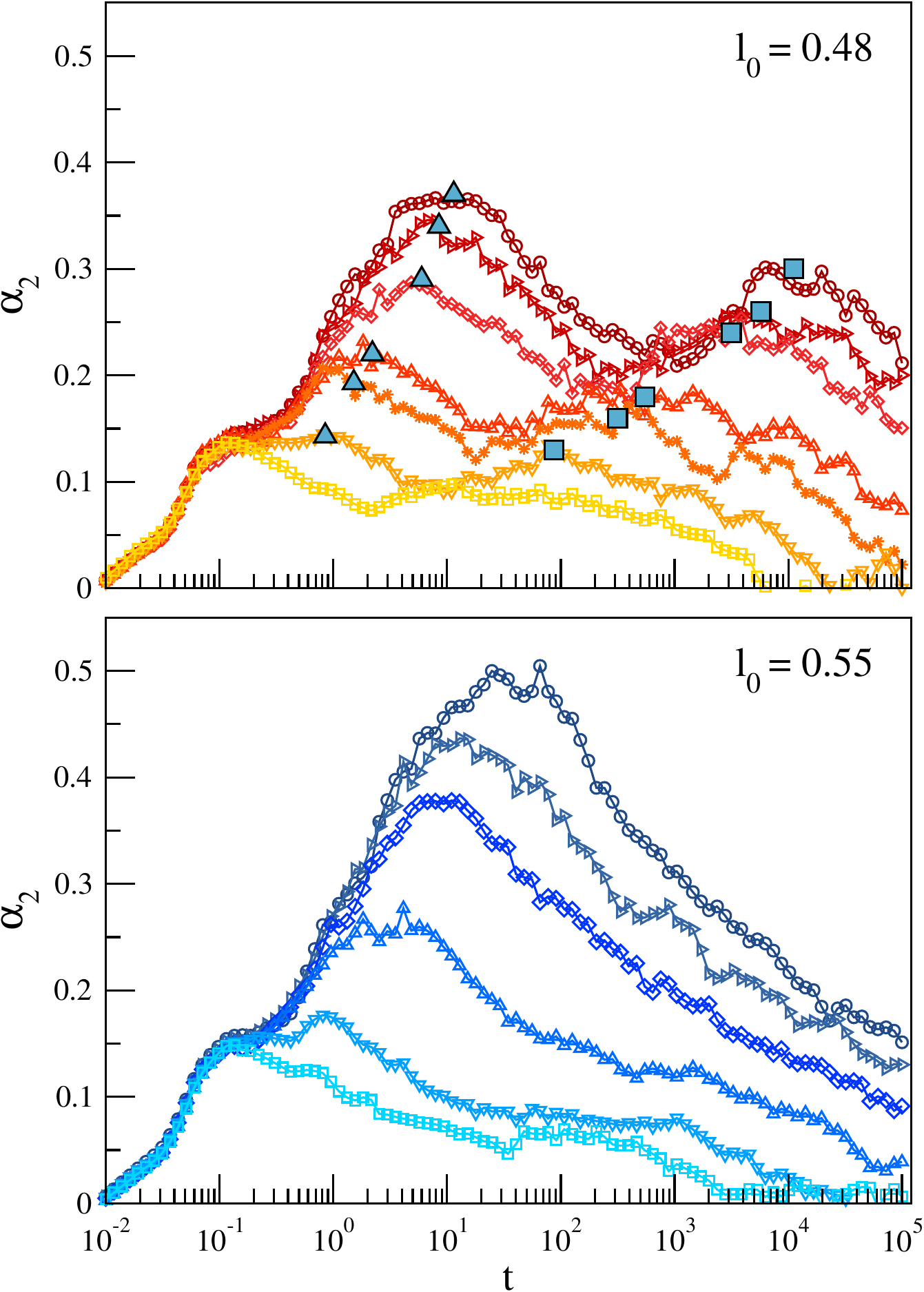}
	\caption{NGP with bond length $\ell_0 = 0.48$ (top) and $\ell_0 = 0.55$ (bottom)  at different temperatures. Color codes as in Fig.\ref{ct}. In the presence  of JG relaxation ($\ell_0 = 0.48$) two peaks increasingly grow by lowering $T$ with positions close to the knees observed in BCF associated to the $\beta$ and the $\alpha$ relaxations, see Fig.\ref{ct}. }
	\label{alpha2}
\end{figure}

\subsection{Dynamic heterogeneity}
\label{DH_NGP}

Fig.\ref{alpha2} compares the time evolution of DH of the melts of chains with different bond length by resorting to their NGP. 
At an early stage ($t \lesssim 0.1$), { the ballistic regime is interrupted by the first repeated collisions of the monomers with the surrounding cage and the subsequent} NGP growth is largely independent of both the temperature and the bond length. The missing role played by the temperature suggests that DH is driven by static structure at short times, as also concluded by previous works  \cite{FragiadakisRolandAlfaBetaPRE17}. The small peak at $t \sim 0.1$ locates the average time needed by the monomer to hit the cage of the first neighbours \cite{CristianoSE}.
For $t  > 0.1$, DH is fully developed and NGP exhibits a complex pattern, strongly dependent on both temperature and bond length.   

\begin{figure}
	\centering
	\includegraphics[width=0.60\linewidth]{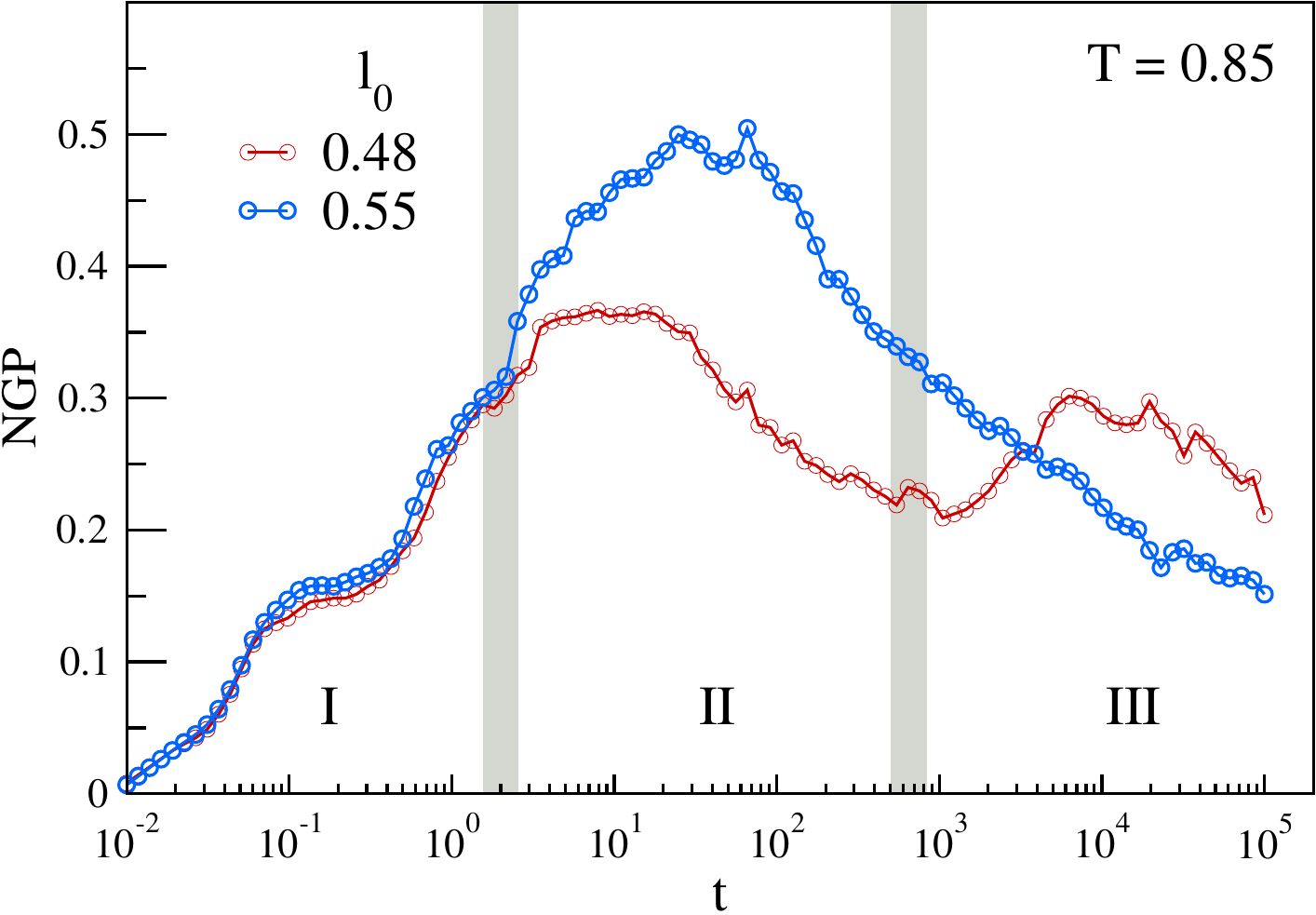}
	\caption{Time evolution of NGP at $T=0.85$ for the two bond lengths.}
	\label{peaks_pos}
\end{figure}

To reach better insight, Fig.\ref{peaks_pos} compares the NGP time evolution of the chains with the two bond lengths at the same temperature (the lowest investigated). Three regions are seen. 
\begin{itemize}
\item Region I: at short times ($t \lesssim 2$) the NGPs are nearly coincident. 
\item Region II: at intermediate times ($2 \lesssim t \lesssim 7 \cdot 10^2$) in the presence of {\it shorter} bonds, switching the $\beta$ relaxation  on in this region, NGP is {\it lower}, i.e. DHs are {\it weaker}, and, after a maximum, decreases to a local minimum. Chains with {\it longer} bonds, with {\it no}  apparent JG relaxation, do not show the minimum, as already known \cite{NGP_Wolynes}.
\item Region III: at long times ($t \gtrsim 7 \cdot 10^2$), where the $\alpha$ relaxation takes place, in the presence of {\it shorter} bonds, NGP reaches a second maximum and finally decays, whereas the NGP of chains with longer bonds decreases monotonously.
\end{itemize}

\subsection{Correlation between relaxation and dynamic heterogeneity}
\label{correlDH_Relax}

There is a well-defined correlation between the DH evolution and relaxation.
To deepen this aspect, in addition to BCF, dealing with  {\it bond reorientation}, we also consider 
the correlation loss of the torsional angle (TACF) \cite{BedrovPRE2005,BedrovJNCS2011}, see SI for rigorous definition. We also inspect quantities concerning the {\it monomer dynamics}: (i) the mean square displacement (MSD) $\langle  \delta r^2 (t) \rangle$ where $\delta r^2 (t)$ is the square modulus of the monomer displacement, $\delta {\bf r}(t)$,  in a time $t$, (ii) the self-part of the intermediate scattering function (ISF) $F_s(q,t) = \langle \exp[ i {\bf q} \cdot \delta {\bf r}(t)] \rangle$ where $i^2=-1$, $q$ is the modulus of the wavevector ${\bf q}$. ISF is negligibly small if the displacement exceeds the length scale $2 \pi/ q$. We choose $q=q_{max}$, where the static structure factor is maximum, so that $2 \pi/ q_{max} \sim \sigma^\star$, i.e. about the monomer diameter.  We stress that, differently from TACF and BCF, both MSD and ISF are {\it single-particle} observables.

\begin{figure}
	\centering
	\includegraphics[width=0.70\linewidth]{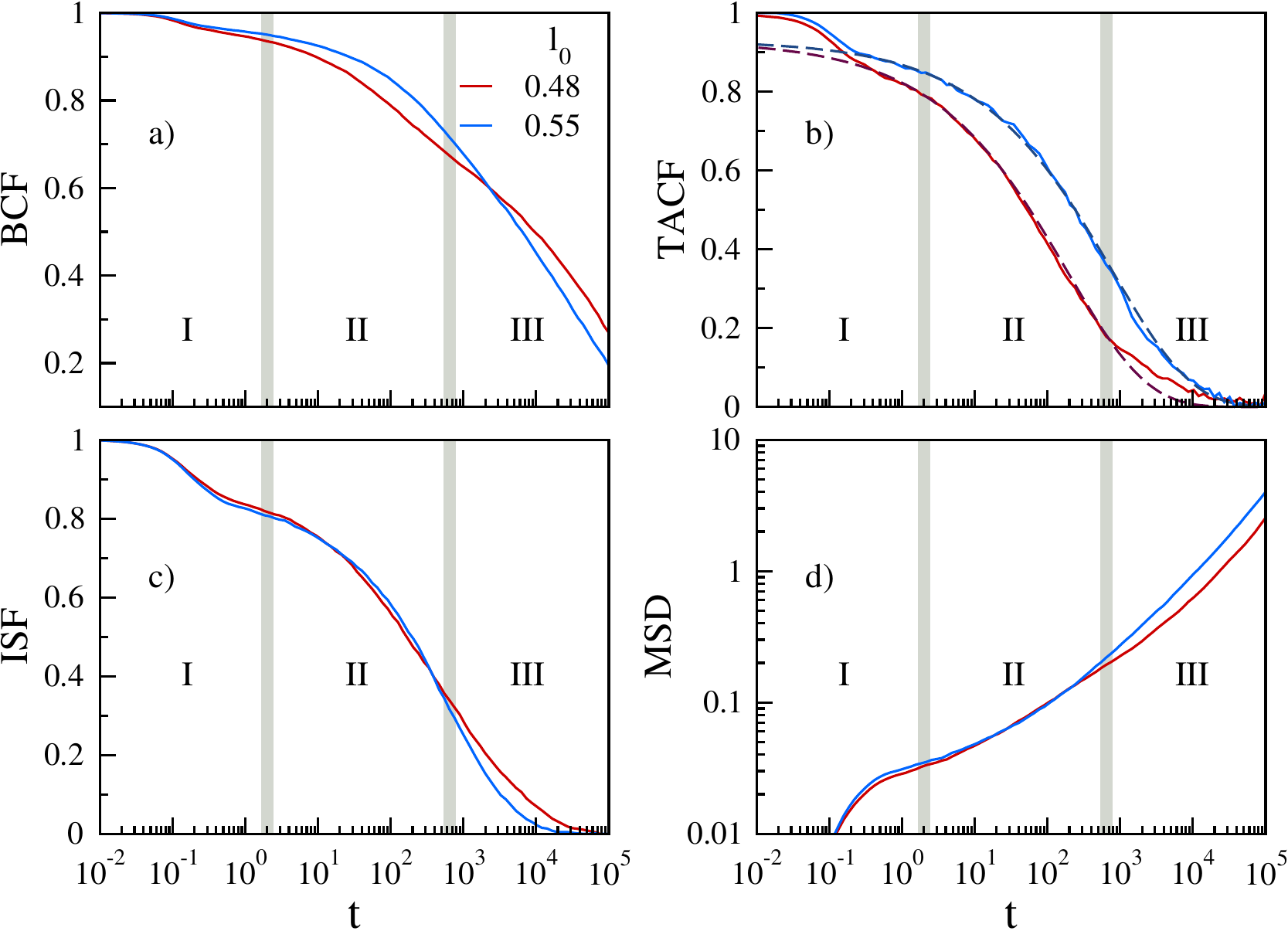}
	\caption{from top left to bottom right: BCF, TACF, ISF and MSD at $T=0.85$ for the two bond lengths. Panels are partitioned in the same three regions of Fig. \ref{peaks_pos}. The dashed curves in the TACF panel are best fits with the stretched function $A \exp[-(t/\tau_{\ell_0})^\beta]$ with $A=0.93$ , $\beta=0.4$, $\tau_{0.48} = 190$ (red dashed), $\tau_{0.55}= 797$ (blue dashed).}
	\label{peaks_pos085}
\end{figure}

Fig.\ref{peaks_pos085} presents the angular relaxation functions (BCF and TACF) together with monomer MSD and ISF, for the two bond lengths at the lowest temperature. Both BCF and TACF clearly show a larger decay in region I and II if the bond is {\it shorter}, i.e. the latter undergoes  {\it faster} small-angle reorientation at short times, especially in the $\beta$ region. In spite of the higher angular mobility at short times,  both BCF and TACF  signal that {\it shorter} bonds {\it slow down} in the $\alpha$ region (region III).  While the reduced decay rate is apparent in BCF, to better visualise the effect in TACF, we first fitted the decay of TACF of chains with longer bonds with a stretched decay in regions II and III which, then, was log-shifted and superimposed to the TACF of chains with shorter bonds. We argue that the slowing down is due to the torsional hindrance due to shorter bonds, see Fig.\ref{bond}b.
The accelerated reorientation of shorter bonds  at short times and their slowing down at  long times have strong influence on the pattern of NGP.  
In fact, the crossover between regions I and II of NGP, see Fig.\ref{peaks_pos}, occurs when the reorientation of shorter bonds becomes apparent, see BCF in Fig.\ref{peaks_pos085}. This suggests that the reduction of NGP observed in polymer systems with  shorter bonds follows by a  partial averaging of  DH.
The latter effect is due to the fast bond  reorientation which inhibits the DH increase, finally resulting in the bump observed in NGP at $t \sim 10$. Naturally, the complete DH erasure needs wide changes in the chain conformation and then full dihedral torsions. This explains why NGP, like BCF, decays more slowly in region III in the presence of shorter bonds. In particular, the peak of NGP in region III for shorter bonds is interpreted as due to the  large scale structural relaxation triggered by the wide-angle torsions.  
{ Finally, by inspecting  both  ISF and MSD in Fig.\ref{peaks_pos085} one sees that the chains with {\it shorter} bonds exhibit {\it smaller} monomer MSD in regions I and III, and {\it slower} structural relaxation in region III, as previously reported \cite{TripodoBetaPolymers2020}.} { The role of ISF to model the memory kernel in a generalized Langevin equation theory dealing with both primary and JG relaxations has been emphasized 
\cite{ZacconeBetaPRB18,ZacconeTamaritBetaPRE18,ZacconeTamaritBetaPRB2020,Zaccone_ReviewRelaxVibratGlasses_JPCM2020}.}

The finding that,  in the presence of JG relaxation,  DH built up in the JG regime persists up to the $\alpha$ relaxation, see Fig.\ref{peaks_pos}, conveys the impression of the possible interplay of DHs in the JG and the long-time regimes, also on the basis that long-time DH is sensed at short times \cite{Harrowell06}, including JG time scale for non-polymeric systems \cite{KarmakarSastryCooperativeBetaPRL16}, and even at vibrational time scales \cite{OurNatPhys,CiceroneDynHeterogJohariGoldsteinPRL14,ReviewIJMS}. 

\subsection{Negligible memory between particle displacements occurring in JG and $\alpha$ time scales}
\label{memory}

The sound assessment of the previous hinted correlation must consider if it holds not only on the {\it whole} system but on {\it subsets} too \cite{BohmerAlfaBetaPRL06_Cit85}.  In non-polymeric liquids NMR experiments answered affirmatively by selecting subensembles of particles with given mobility \cite{BohmerAlfaBetaPRL06_Cit85},  whereas simulations on diatomic liquids concluded that there is no connection between properties associated to beta and alpha regimes of a single molecule \cite{FragiadakisRolandAlfaBetaPRE17}. { On the other hand, the invariance of the ratio $\tau_{JG}/\tau_\alpha$ to $P$ and $T$ variations was reported in linear polymer melts represented by a simple bead-necklace model \cite{BedrovJNCS2011}, a variant of which is adopted in the present paper.
}
\begin{figure}
	\centering
	\includegraphics[width=0.40\linewidth]{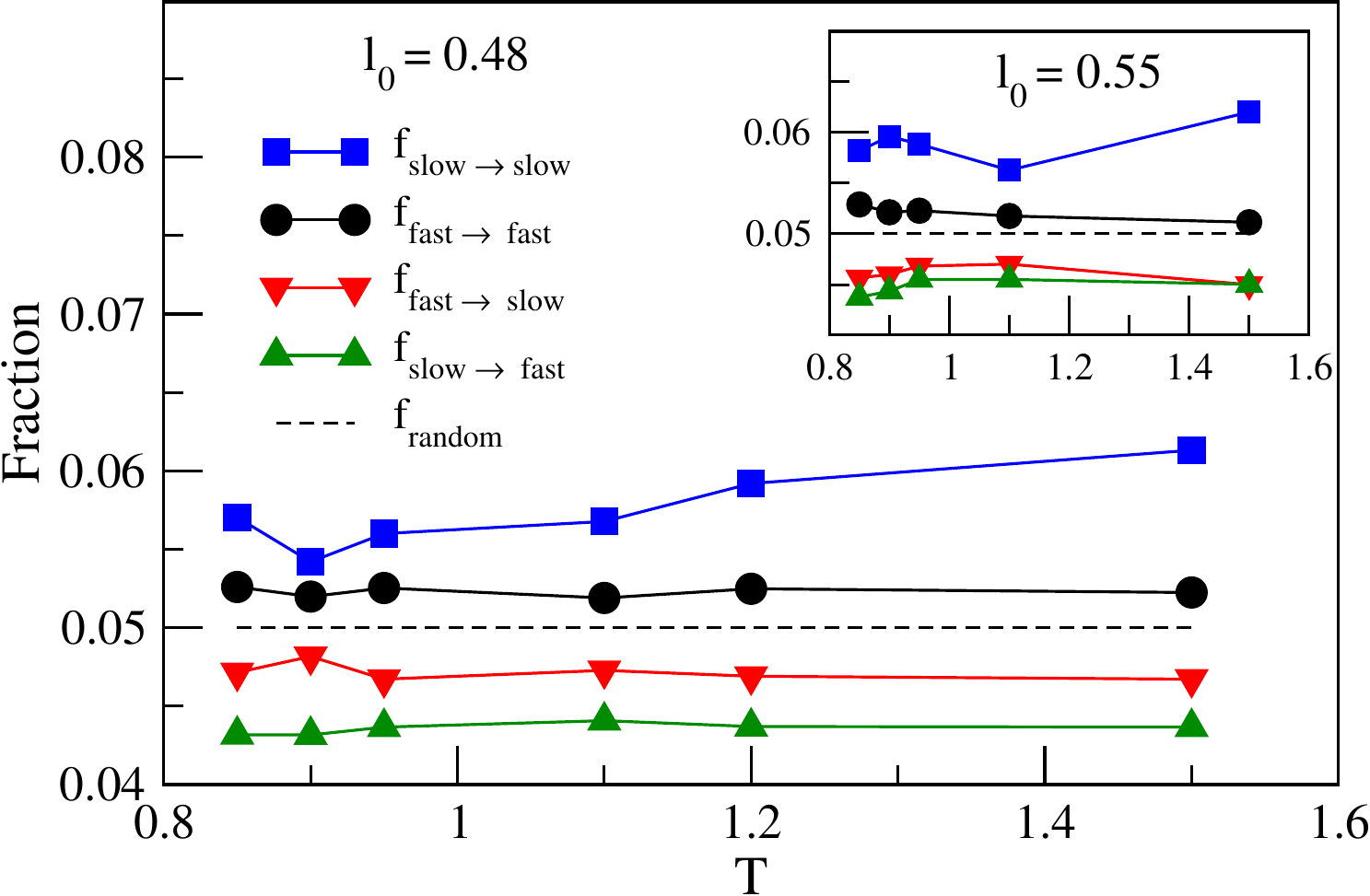}
	\caption{Fraction of monomers of polymer system with JG relaxation ($l_0=0.48$) retaining memory of their slow (or fast) mobility at JG time scale up to primary relaxation, $f_{slow \to slow}$ and $f_{fast \to fast}$ respectively. In the case of full memory  
	$f_{slow \to slow}=f_{fast \to fast}=1$. Dashed line sets the level $f_{random}=0.05$ corresponding to the absence of memory. Cross-conversion fractions are also plotted. The results point to very weak memory.  See text for details. Inset: same analysis with $l_0=0.55$.}
	\label{fraction}
\end{figure}
We investigate this aspect by considering the system with JG relaxation ($l_0=0.48$) and focusing on the DH developed at $t_\beta$, the time of the first peak of NGP (located in the $\beta$ region) and DH still surviving at $t_\alpha$, the time of the second peak (located in the $\alpha$ region), see fig.\ref{peaks_pos}. We search for "memory" effects between $t_\beta$ and $t_\alpha$ by: i) selecting a subset of $n$ monomers ($n = 0.05 N$, $N$ being the total monomers) with smallest (or largest) displacement in a time lapse $t_\beta$, and ii) evaluating the fraction $f_{slow \to slow}$ (or $f_{fast \to fast}$) of these monomers still belonging to the same mobility subset of $n$ monomers with lowest (or highest) mobility after displacement in a time $t_\alpha$ (in the case of {\it full} memory  $f_{slow \to slow}=f_{fast \to fast}=1$, whereas in the {\it absence} of memory $f_{slow \to slow}=f_{fast \to fast}=f_{random}$ with $f_{random} = n/N$, with the latter result following if the initial subset in step i)  is assembled by picking-up the $n$  monomers {\it randomly}). 
Fig.\ref{fraction} shows the results. It is seen that both $f_{slow \to slow}$ and $f_{fast \to fast}$ are only slightly larger than $f_{random}$, i.e. the correlation between the two subsets is quite low. The same conclusion is reached by considering the fractions $f_{slow \to fast}$ and $f_{fast \to slow}$ accounting for the cross-conversion between the two mobility subsets (note that in the case of full memory both quantities {\it vanish}). The pattern does not change by considering {\it longer} bonds, i.e. no JG relaxation, and different ratios $n/N$ ($n/N=0.03,0.1$, not shown).

\section{Conclusions}
\label{res_disc}

In conclusion, we evidenced the heterogeneous character of the JG relaxation by MD simulations of a model polymer with secondary relaxation tuned by an effective adjustable torsional barrier. The DH evolution exhibits two distinct maxima at JG and structural relaxation time scales. We find that subsets of monomers lose memory of their mobility acquired in JG time scale  before the occurrence of structural relaxation.

%%%%%%%%%%%%%%%%%%%%%%%%%%%%%%%%%%%%%%%%%%%%%%%%%%%%%%%%%%%%%%%%%%%%%
%% The "Acknowledgement" section can be given in all manuscript
%% classes.  This should be given within the "acknowledgement"
%% environment, which will make the correct section or running title.
%%%%%%%%%%%%%%%%%%%%%%%%%%%%%%%%%%%%%%%%%%%%%%%%%%%%%%%%%%%%%%%%%%%%%
\begin{acknowledgement}
Simone Capaccioli is warmly thanked for discussions. We acknowledge the support from the project PRA-2018-34 ("ANISE") from the University of Pisa. A generous grant of computing time from IT Center, University of Pisa and Dell EMC${}^\circledR$ Italia is also gratefully acknowledged. 
\end{acknowledgement}

%%%%%%%%%%%%%%%%%%%%%%%%%%%%%%%%%%%%%%%%%%%%%%%%%%%%%%%%%%%%%%%%%%%%%
%% The same is true for Supporting Information, which should use the
%% suppinfo environment.
%%%%%%%%%%%%%%%%%%%%%%%%%%%%%%%%%%%%%%%%%%%%%%%%%%%%%%%%%%%%%%%%%%%%%
\begin{suppinfo}

\subsection*{Simulation Details}

\noindent
We study a melt of coarse-grained linear polymer chains with $N_c = 512$ linear chains made of $M=25$ monomers each, resulting in a total number of monomers $N=12800$. 
Adjacent bonded monomers belonging to the same chain  interact via the harmonic potential 
	$U^{bond}(r)=k_{bond}\left(r-l_0\right)^2$, where the constant $k_{bond}$ is set to $2000 \epsilon/\sigma^2$ to ensure high stiffness. A bending potential $U^{bend}(\alpha)=k_{bend}\left(\cos \alpha-\cos\alpha_0\right)^2$, with $k_{bend}=2000 \epsilon$ and $\alpha_0=120^\circ$, is introduced to maintain the angle $\alpha$ between two consecutive bonds nearly constant.
Non-adjacent monomers in the same chain or monomers belonging to different chains are defined as "non-bonded" monomers. Non-bonded monomers, when placed at mutual distance $r$, interact via a shifted Lennard-Jones (LJ) potential:
	\begin{equation}
	U^{LJ}(r) = {\epsilon} \left [   \left (\frac{\sigma^*}{r}\right )^{12} - 2 \left (\frac{\sigma^*}{r}\right )^{6}\right ] + U_{cut}
	\label{Eq:modifiedLJ},
	\end{equation}
	where $\sigma^* = 2^{1/6} \sigma$ is the minimum of the potential,  $U^{LJ}(r = \sigma^*) = -\epsilon + U_{cut}$. The potential is truncated at $r=r_c=2.5\sigma$ for computational convenience and the constant $U_{cut}$ adjusted to ensure that $U^{LJ}(r)$ is continuous at  $r=r_c$ with $U^{LJ}(r)=0$ for $r \ge r_c$. 
	
	It is important to note that the above model allows LJ interactions between {\it all} non-bonded monomers. This is the feature to build the torsional barrier up when $l_0 < 0.5  \sigma$ discussed in the paper. Alternatives reported in the literature exclude the LJ interactions between atoms separated by three bonds or less \cite{BedrovJNCS2011}.
		
	All the data presented in the work are expressed in reduced MD units: length in units of $\sigma$, temperature in units of $\epsilon/k_B$, where $k_B$ is the Boltzmann constant, and time in units of $\tau_{MD} = (m\sigma^2 / \epsilon)^{1/2}$. We set $\sigma = 1$, $\epsilon = 1$,  $m = 1$ and $k_B = 1$. 
	
	Simulations were carried out with the open-source software LAMMPS \cite{PlimptonLAMMPS,PlimptonURL}. Equilibration runs were performed at constant number of monomers $N$, constant vanishing pressure $P=0$ and constant temperature $T$ ($NPT$ ensemble). For each state the equilibration lasted at least for $3 \tau_{ee}$, being $\tau_{ee}$  the relaxation time of the end-to-end vector autocorrelation function \cite{DoiEdwards}. Production runs have been performed within the $NVT$ ensemble (constant number of monomers $N$, constant volume $V$ and constant temperature $T$). Additional short equilibration runs were performed when switching from $NPT$ to $NVT$ ensemble. No signatures of crystallization were observed in all the investigated states.  
	
%\newpage

\subsection*{Torsional correlation function}
Alternatively to the BCF, the relaxation of the chain backbone arrangement has been characterized by the torsional autocorrelation function (TACF) \cite{BedrovJNCS2011}:
\begin{equation}
TACF(t)=\frac{1}{N_c}\frac{1}{M-3}\sum_{n=1}^{N_c}\sum_{m=1}^{M-3}\frac{\langle|\varphi_{m,n}(t)||\varphi_{m,n}(0)|\rangle-\langle|\varphi_{m,n}(0)|\rangle^2}{\langle|\varphi_{m,n}(0)|^2\rangle-\langle|\varphi_{m,n}(0)|\rangle^2},
\label{TACF}
\end{equation}
where $|\varphi_{m,n}(t)|$ is the modulus of the $m$-th dihedral angle of the $n$-th chain at a given time $t$ \cite{BedrovJNCS2011,KremerGrestCarmesinPRL88}. $\varphi$ is trivially related to the angle $\phi$ considered in Fig. 1 of the manuscript by the relation $\varphi=\phi-180^\circ$.
The dihedral angle features the torsion of a given bond. Terminal bonds are not subject to torsion, therefore, in a chain of length $M$, there are $M-3$ dihedral angles. A graphical representation of this angle is reported in Fig.\ref{dihedral_angle}: $\varphi$ is given from the intersection of the two planes defined by four consecutive monomers in a chain: the first plane is defined considering the first set of three monomers while the second one is defined by the last set of three. 
\begin{figure}[t]
	\centering
	\includegraphics[width=0.5\textwidth]{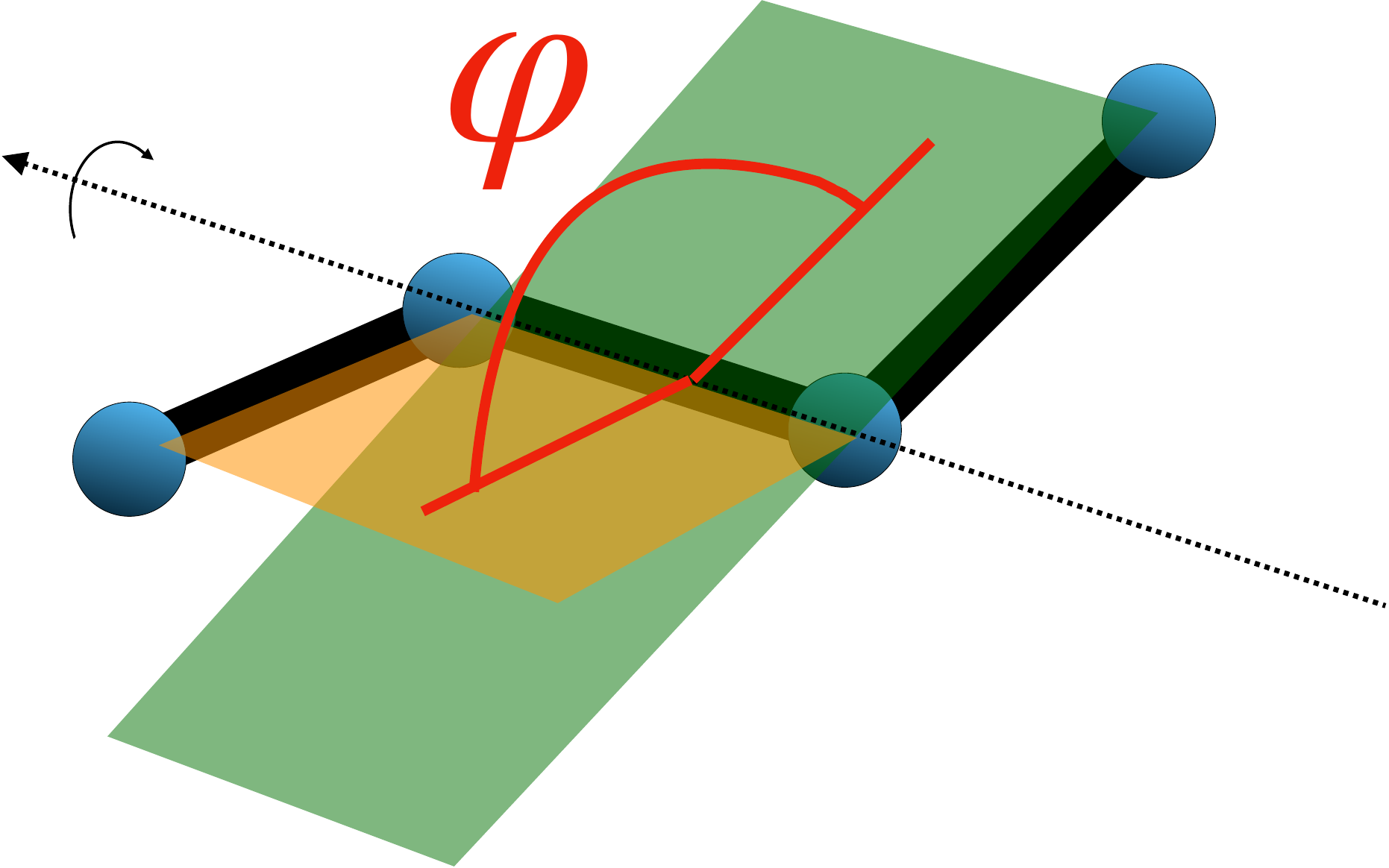}
	\caption{Representation of the dihedral angle, $\varphi$, characterizing the torsion of the central bond. }
	\label{dihedral_angle}
\end{figure}

\begin{figure}[t]
	\centering
	\includegraphics[width=\linewidth]{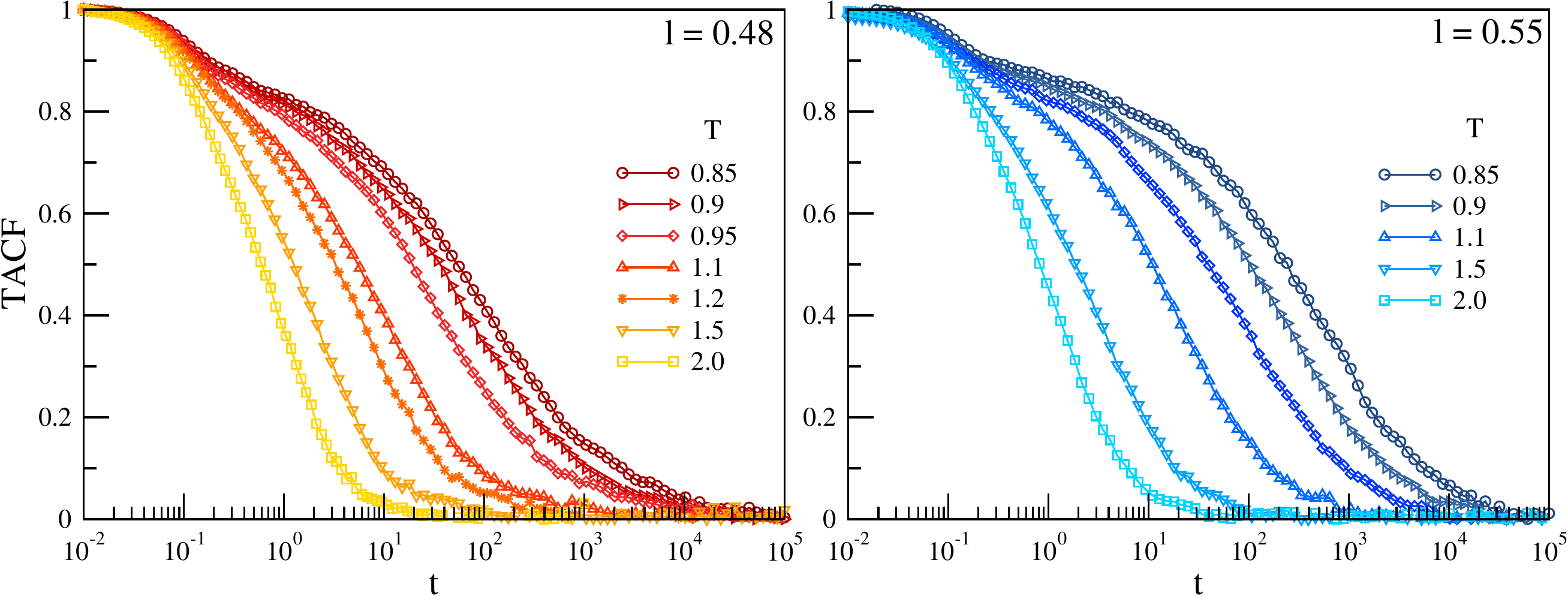}
	\caption{Temperature dependence of TACF in the two studied systems.}
	\label{tacf}
\end{figure}

The temperature dependence of the TACF curves is reported in Fig.\ref{tacf}.

\end{suppinfo}

%%%%%%%%%%%%%%%%%%%%%%%%%%%%%%%%%%%%%%%%%%%%%%%%%%%%%%%%%%%%%%%%%%%%%
%% The appropriate \bibliography command should be placed here.
%% Notice that the class file automatically sets \bibliographystyle
%% and also names the section correctly.
%%%%%%%%%%%%%%%%%%%%%%%%%%%%%%%%%%%%%%%%%%%%%%%%%%%%%%%%%%%%%%%%%%%%%

\newpage

\bibliography{biblio}

\providecommand{\latin}[1]{#1}
\makeatletter
\providecommand{\doi}
  {\begingroup\let\do\@makeother\dospecials
  \catcode`\{=1 \catcode`\}=2 \doi@aux}
\providecommand{\doi@aux}[1]{\endgroup\texttt{#1}}
\makeatother
\providecommand*\mcitethebibliography{\thebibliography}
\csname @ifundefined\endcsname{endmcitethebibliography}
  {\let\endmcitethebibliography\endthebibliography}{}
\begin{mcitethebibliography}{61}
\providecommand*\natexlab[1]{#1}
\providecommand*\mciteSetBstSublistMode[1]{}
\providecommand*\mciteSetBstMaxWidthForm[2]{}
\providecommand*\mciteBstWouldAddEndPuncttrue
  {\def\EndOfBibitem{\unskip.}}
\providecommand*\mciteBstWouldAddEndPunctfalse
  {\let\EndOfBibitem\relax}
\providecommand*\mciteSetBstMidEndSepPunct[3]{}
\providecommand*\mciteSetBstSublistLabelBeginEnd[3]{}
\providecommand*\EndOfBibitem{}
\mciteSetBstSublistMode{f}
\mciteSetBstMaxWidthForm{subitem}{(\alph{mcitesubitemcount})}
\mciteSetBstSublistLabelBeginEnd
  {\mcitemaxwidthsubitemform\space}
  {\relax}
  {\relax}

\bibitem[Debenedetti(1997)]{DebenedettiBook}
Debenedetti,~P.~G. \emph{Metastable Liquids}; Princeton University Press,
  Princeton USA, 1997\relax
\mciteBstWouldAddEndPuncttrue
\mciteSetBstMidEndSepPunct{\mcitedefaultmidpunct}
{\mcitedefaultendpunct}{\mcitedefaultseppunct}\relax
\EndOfBibitem
\bibitem[McCrum \latin{et~al.}(1991)McCrum, Read, and Williams]{WilliamsMcCrum}
McCrum,~N.~G.; Read,~B.~E.; Williams,~G. \emph{Anelastic and Dielectric Effects
  in Polymeric Solids}; Dover Publications: New York, 1991\relax
\mciteBstWouldAddEndPuncttrue
\mciteSetBstMidEndSepPunct{\mcitedefaultmidpunct}
{\mcitedefaultendpunct}{\mcitedefaultseppunct}\relax
\EndOfBibitem
\bibitem[Angell \latin{et~al.}(2000)Angell, Ngai, McKenna, McMillan, and
  S.W.Martin]{AngelNgai00}
Angell,~C.~A.; Ngai,~K.~L.; McKenna,~G.~B.; McMillan,~P.; S.W.Martin,
  Relaxation in glassforming liquids and amorphous solids. \emph{J. Appl.
  Phys.} \textbf{2000}, \emph{88}, 3113--3157\relax
\mciteBstWouldAddEndPuncttrue
\mciteSetBstMidEndSepPunct{\mcitedefaultmidpunct}
{\mcitedefaultendpunct}{\mcitedefaultseppunct}\relax
\EndOfBibitem
\bibitem[Ngai(2011)]{NgaiBook}
Ngai,~K.~L. \emph{Relaxation and Diffusion in Complex Systems}; Springer,
  Berlin, 2011\relax
\mciteBstWouldAddEndPuncttrue
\mciteSetBstMidEndSepPunct{\mcitedefaultmidpunct}
{\mcitedefaultendpunct}{\mcitedefaultseppunct}\relax
\EndOfBibitem
\bibitem[Johari and Goldstein(1970)Johari, and Goldstein]{JOHARI70}
Johari,~G.~P.; Goldstein,~M. Viscous Liquids and the Glass Transition. II.
  Secondary Relaxations in Glasses of Rigid Molecules. \emph{J. Chem. Phys.}
  \textbf{1970}, \emph{53}, 2372\relax
\mciteBstWouldAddEndPuncttrue
\mciteSetBstMidEndSepPunct{\mcitedefaultmidpunct}
{\mcitedefaultendpunct}{\mcitedefaultseppunct}\relax
\EndOfBibitem
\bibitem[Ngai(1998)]{Ngai98}
Ngai,~K. Relation between some secondary relaxations and the a relaxations in
  glass-forming materials according to the coupling model. \emph{J. Chem.
  Phys.} \textbf{1998}, \emph{109}, 6982--6994\relax
\mciteBstWouldAddEndPuncttrue
\mciteSetBstMidEndSepPunct{\mcitedefaultmidpunct}
{\mcitedefaultendpunct}{\mcitedefaultseppunct}\relax
\EndOfBibitem
\bibitem[Ngai and Paluch(2004)Ngai, and
  Paluch]{NgaiPaluchClassificationSecondaryJCP04}
Ngai,~K.~L.; Paluch,~M. Classification of secondary relaxation in glass-formers
  based on dynamic properties. \emph{The Journal of Chemical Physics}
  \textbf{2004}, \emph{120}, 857--873\relax
\mciteBstWouldAddEndPuncttrue
\mciteSetBstMidEndSepPunct{\mcitedefaultmidpunct}
{\mcitedefaultendpunct}{\mcitedefaultseppunct}\relax
\EndOfBibitem
\bibitem[Capaccioli \latin{et~al.}(2012)Capaccioli, Paluch, Prevosto, Wang, and
  Ngai]{Capaccioli12}
Capaccioli,~S.; Paluch,~M.; Prevosto,~D.; Wang,~L.-M.; Ngai,~K.~L. Many-Body
  Nature of Relaxation Processes in Glass-Forming Systems. \emph{The Journal of
  Physical Chemistry Letters} \textbf{2012}, \emph{3}, 735--743\relax
\mciteBstWouldAddEndPuncttrue
\mciteSetBstMidEndSepPunct{\mcitedefaultmidpunct}
{\mcitedefaultendpunct}{\mcitedefaultseppunct}\relax
\EndOfBibitem
\bibitem[Cicerone \latin{et~al.}(2014)Cicerone, Zhong, and
  Tyagi]{CiceroneDynHeterogJohariGoldsteinPRL14}
Cicerone,~M.~T.; Zhong,~Q.; Tyagi,~M. Picosecond Dynamic Heterogeneity,
  Hopping, and Johari-Goldstein Relaxation in Glass-Forming Liquids.
  \emph{Phys. Rev. Lett.} \textbf{2014}, \emph{113}, 117801, DOI:
  \doi{10.1103/PhysRevLett.113.117801}\relax
\mciteBstWouldAddEndPuncttrue
\mciteSetBstMidEndSepPunct{\mcitedefaultmidpunct}
{\mcitedefaultendpunct}{\mcitedefaultseppunct}\relax
\EndOfBibitem
\bibitem[H.~B.~Yu and Samwer(2017)H.~B.~Yu, and Samwer]{yu2017}
H.~B.~Yu,~R.~R.; Samwer,~K. Structural rearrangements governing
  Johari-Goldstein relaxations in metallic glasses. \emph{Sci. Adv.}
  \textbf{2017}, \emph{3}, 1701577\relax
\mciteBstWouldAddEndPuncttrue
\mciteSetBstMidEndSepPunct{\mcitedefaultmidpunct}
{\mcitedefaultendpunct}{\mcitedefaultseppunct}\relax
\EndOfBibitem
\bibitem[Boyd and Breitling(1974)Boyd, and Breitling]{Boyd:1974lq}
Boyd,~R.~H.; Breitling,~S.~M. The Conformational Analysis of Crankshaft Motions
  in Polyethylene. \emph{Macromolecules} \textbf{1974}, \emph{7}, 855--862,
  DOI: \doi{10.1021/ma60042a032}\relax
\mciteBstWouldAddEndPuncttrue
\mciteSetBstMidEndSepPunct{\mcitedefaultmidpunct}
{\mcitedefaultendpunct}{\mcitedefaultseppunct}\relax
\EndOfBibitem
\bibitem[Paul \latin{et~al.}(1997)Paul, Smith, and Yoon]{Paul:1997wd}
Paul,~W.; Smith,~G.~D.; Yoon,~D.~Y. Static and Dynamic Properties of a
  n-C100H202 Melt from Molecular Dynamics Simulations. \emph{Macromolecules}
  \textbf{1997}, \emph{30}, 7772--7780, DOI: \doi{10.1021/ma971184d}\relax
\mciteBstWouldAddEndPuncttrue
\mciteSetBstMidEndSepPunct{\mcitedefaultmidpunct}
{\mcitedefaultendpunct}{\mcitedefaultseppunct}\relax
\EndOfBibitem
\bibitem[Meier and Struik(1998)Meier, and Struik]{Struik98}
Meier,~R.~J.; Struik,~L. Atomistic modelling study of relaxation processes in
  polymers: the $\beta$-relaxation in polyvinylchloride. \emph{Polymer}
  \textbf{1998}, \emph{39}, 31--38\relax
\mciteBstWouldAddEndPuncttrue
\mciteSetBstMidEndSepPunct{\mcitedefaultmidpunct}
{\mcitedefaultendpunct}{\mcitedefaultseppunct}\relax
\EndOfBibitem
\bibitem[Goldstein(2011)]{GOLDSTEIN_JNCS11}
Goldstein,~M. The past, present, and future of the Johari--Goldstein
  relaxation. \emph{Journal of Non-Crystalline Solids} \textbf{2011},
  \emph{357}, 249 -- 250, DOI:
  \doi{https://doi.org/10.1016/j.jnoncrysol.2010.05.105}\relax
\mciteBstWouldAddEndPuncttrue
\mciteSetBstMidEndSepPunct{\mcitedefaultmidpunct}
{\mcitedefaultendpunct}{\mcitedefaultseppunct}\relax
\EndOfBibitem
\bibitem[Johari(1976)]{Johari76}
Johari,~G.~P. GLASS TRANSITION AND SECONDARY RELAXATIONS IN MOLECULAR LIQUIDS
  AND CRYSTALS. \emph{Annals of the New York Academy of Sciences}
  \textbf{1976}, \emph{279}, 117--140, DOI:
  \doi{10.1111/j.1749-6632.1976.tb39701.x}\relax
\mciteBstWouldAddEndPuncttrue
\mciteSetBstMidEndSepPunct{\mcitedefaultmidpunct}
{\mcitedefaultendpunct}{\mcitedefaultseppunct}\relax
\EndOfBibitem
\bibitem[Bershtein \latin{et~al.}(1994)Bershtein, Egorov, Egorova, and
  Ryzhov]{BERSHTEIN199441}
Bershtein,~V.; Egorov,~V.; Egorova,~L.; Ryzhov,~V. The role of thermal analysis
  in revealing the common molecular nature of transitions in polymers.
  \emph{Thermochimica Acta} \textbf{1994}, \emph{238}, 41 -- 73, DOI:
  \doi{https://doi.org/10.1016/S0040-6031(94)85206-5}\relax
\mciteBstWouldAddEndPuncttrue
\mciteSetBstMidEndSepPunct{\mcitedefaultmidpunct}
{\mcitedefaultendpunct}{\mcitedefaultseppunct}\relax
\EndOfBibitem
\bibitem[Ngai(1998)]{NgaiAlfaBetaCor98}
Ngai,~K.~L. Relation between some secondary relaxations and the $\alpha$
  relaxations in glass-forming materials according to the coupling model.
  \emph{The Journal of Chemical Physics} \textbf{1998}, \emph{109}, 6982--6994,
  DOI: \doi{10.1063/1.477334}\relax
\mciteBstWouldAddEndPuncttrue
\mciteSetBstMidEndSepPunct{\mcitedefaultmidpunct}
{\mcitedefaultendpunct}{\mcitedefaultseppunct}\relax
\EndOfBibitem
\bibitem[B\"ohmer \latin{et~al.}(2006)B\"ohmer, Diezemann, Geil, Hinze,
  Nowaczyk, and Winterlich]{BohmerAlfaBetaPRL06_Cit85}
B\"ohmer,~R.; Diezemann,~G.; Geil,~B.; Hinze,~G.; Nowaczyk,~A.; Winterlich,~M.
  Correlation of Primary and Secondary Relaxations in a Supercooled Liquid.
  \emph{Phys. Rev. Lett.} \textbf{2006}, \emph{97}, 135701, DOI:
  \doi{10.1103/PhysRevLett.97.135701}\relax
\mciteBstWouldAddEndPuncttrue
\mciteSetBstMidEndSepPunct{\mcitedefaultmidpunct}
{\mcitedefaultendpunct}{\mcitedefaultseppunct}\relax
\EndOfBibitem
\bibitem[Goldstein(2010)]{GoldsteinBetaDynbamicalHetJCP10}
Goldstein,~M. Communications: Comparison of activation barriers for the
  Johari--Goldstein and alpha relaxations and its implications. \emph{The
  Journal of Chemical Physics} \textbf{2010}, \emph{132}, 041104, DOI:
  \doi{10.1063/1.3306562}\relax
\mciteBstWouldAddEndPuncttrue
\mciteSetBstMidEndSepPunct{\mcitedefaultmidpunct}
{\mcitedefaultendpunct}{\mcitedefaultseppunct}\relax
\EndOfBibitem
\bibitem[Cicerone and Tyagi(2017)Cicerone, and
  Tyagi]{CiceroneMetabasinJG_JCP17}
Cicerone,~M.~T.; Tyagi,~M. Metabasin transitions are Johari-Goldstein
  relaxation events. \emph{The Journal of Chemical Physics} \textbf{2017},
  \emph{146}, 054502\relax
\mciteBstWouldAddEndPuncttrue
\mciteSetBstMidEndSepPunct{\mcitedefaultmidpunct}
{\mcitedefaultendpunct}{\mcitedefaultseppunct}\relax
\EndOfBibitem
\bibitem[Charbonneau \latin{et~al.}(2014)Charbonneau, Kurchan, Parisi, Urbani,
  and Zamponi]{CharbonneauKurchanParisiZamponoNatComm14}
Charbonneau,~P.; Kurchan,~J.; Parisi,~G.; Urbani,~P.; Zamponi,~F. Fractal free
  energy landscapes in structural glasses. \emph{Nature Communications}
  \textbf{2014}, \emph{5}, 3725, DOI: \doi{10.1038/ncomms4725}\relax
\mciteBstWouldAddEndPuncttrue
\mciteSetBstMidEndSepPunct{\mcitedefaultmidpunct}
{\mcitedefaultendpunct}{\mcitedefaultseppunct}\relax
\EndOfBibitem
\bibitem[Johari(2019)]{Johari:2019dw}
Johari,~G.~P. Source of JG-Relaxation in the Entropy of Glass. \emph{The
  Journal of Physical Chemistry B} \textbf{2019}, \emph{123}, 3010--3023, DOI:
  \doi{10.1021/acs.jpcb.9b00612}\relax
\mciteBstWouldAddEndPuncttrue
\mciteSetBstMidEndSepPunct{\mcitedefaultmidpunct}
{\mcitedefaultendpunct}{\mcitedefaultseppunct}\relax
\EndOfBibitem
\bibitem[Ngai \latin{et~al.}(2020)Ngai, Capaccioli, Paluch, and
  Wang]{CapaccioliNgaiJGPhilMag20}
Ngai,~K.~L.; Capaccioli,~S.; Paluch,~M.; Wang,~L. Clarifying the nature of the
  Johari-Goldstein $\beta$-relaxation and emphasising its fundamental
  importance. \emph{Philosophical Magazine} \textbf{2020}, \emph{100},
  2596--2613, DOI: \doi{10.1080/14786435.2020.1781276}\relax
\mciteBstWouldAddEndPuncttrue
\mciteSetBstMidEndSepPunct{\mcitedefaultmidpunct}
{\mcitedefaultendpunct}{\mcitedefaultseppunct}\relax
\EndOfBibitem
\bibitem[Wang \latin{et~al.}(2020)Wang, Zhou, Guan, Yu, Wang, and
  Ngai]{NgaiMetallicGlassPRB20}
Wang,~B.; Zhou,~Z.~Y.; Guan,~P.~F.; Yu,~H.~B.; Wang,~W.~H.; Ngai,~K.~L.
  Invariance of the relation between \ensuremath{\alpha} relaxation and
  \ensuremath{\beta} relaxation in metallic glasses to variations of pressure
  and temperature. \emph{Phys. Rev. B} \textbf{2020}, \emph{102}, 094205, DOI:
  \doi{10.1103/PhysRevB.102.094205}\relax
\mciteBstWouldAddEndPuncttrue
\mciteSetBstMidEndSepPunct{\mcitedefaultmidpunct}
{\mcitedefaultendpunct}{\mcitedefaultseppunct}\relax
\EndOfBibitem
\bibitem[Smith and Bedrov(2007)Smith, and Bedrov]{BedrovSmithJPolymSci07}
Smith,~G.~D.; Bedrov,~D. Relationship between the $\alpha$- and
  $\beta$-relaxation processes in amorphous polymers: Insight from atomistic
  molecular dynamics simulations of 1,4-polybutadiene melts and blends.
  \emph{Journal of Polymer Science Part B: Polymer Physics} \textbf{2007},
  \emph{45}, 627--643, DOI: \doi{https://doi.org/10.1002/polb.21064}\relax
\mciteBstWouldAddEndPuncttrue
\mciteSetBstMidEndSepPunct{\mcitedefaultmidpunct}
{\mcitedefaultendpunct}{\mcitedefaultseppunct}\relax
\EndOfBibitem
\bibitem[Karmakar \latin{et~al.}(2016)Karmakar, Dasgupta, and
  Sastry]{KarmakarSastryCooperativeBetaPRL16}
Karmakar,~S.; Dasgupta,~C.; Sastry,~S. Short-Time Beta Relaxation in
  Glass-Forming Liquids Is Cooperative in Nature. \emph{Phys. Rev. Lett.}
  \textbf{2016}, \emph{116}, 085701, DOI:
  \doi{10.1103/PhysRevLett.116.085701}\relax
\mciteBstWouldAddEndPuncttrue
\mciteSetBstMidEndSepPunct{\mcitedefaultmidpunct}
{\mcitedefaultendpunct}{\mcitedefaultseppunct}\relax
\EndOfBibitem
\bibitem[Sillescu(1999)]{SILLESCURevDynHet99}
Sillescu,~H. Heterogeneity at the glass transition: a review. \emph{Journal of
  Non-Crystalline Solids} \textbf{1999}, \emph{243}, 81--108\relax
\mciteBstWouldAddEndPuncttrue
\mciteSetBstMidEndSepPunct{\mcitedefaultmidpunct}
{\mcitedefaultendpunct}{\mcitedefaultseppunct}\relax
\EndOfBibitem
\bibitem[Ediger(2000)]{Ediger00}
Ediger,~M.~D. Spatially heterogeneous dynamics in supercooled liquids.
  \emph{Annu. Rev. Phys. Chem.} \textbf{2000}, \emph{51}, 99--128\relax
\mciteBstWouldAddEndPuncttrue
\mciteSetBstMidEndSepPunct{\mcitedefaultmidpunct}
{\mcitedefaultendpunct}{\mcitedefaultseppunct}\relax
\EndOfBibitem
\bibitem[Richert(2002)]{Richert02}
Richert,~R. Heterogeneous dynamics in liquids: fluctuations in space and time.
  \emph{J. Phys.: Condens. Matter} \textbf{2002}, \emph{14}, R703--R738\relax
\mciteBstWouldAddEndPuncttrue
\mciteSetBstMidEndSepPunct{\mcitedefaultmidpunct}
{\mcitedefaultendpunct}{\mcitedefaultseppunct}\relax
\EndOfBibitem
\bibitem[Berthier and Biroli(2011)Berthier, and Biroli]{BerthieRev}
Berthier,~L.; Biroli,~G. Theoretical perspective on the glass transition and
  amorphous materials. \emph{Rev. Mod. Phys.} \textbf{2011}, \emph{83},
  587--645\relax
\mciteBstWouldAddEndPuncttrue
\mciteSetBstMidEndSepPunct{\mcitedefaultmidpunct}
{\mcitedefaultendpunct}{\mcitedefaultseppunct}\relax
\EndOfBibitem
\bibitem[Karmakar \latin{et~al.}(2015)Karmakar, Dasgupta, and
  Sastry]{SastryLengthScalesRepProgrPhys15}
Karmakar,~S.; Dasgupta,~C.; Sastry,~S. Length scales in glass-forming liquids
  and related systems: a review. \emph{Reports on Progress in Physics}
  \textbf{2015}, \emph{79}, 016601\relax
\mciteBstWouldAddEndPuncttrue
\mciteSetBstMidEndSepPunct{\mcitedefaultmidpunct}
{\mcitedefaultendpunct}{\mcitedefaultseppunct}\relax
\EndOfBibitem
\bibitem[Napolitano \latin{et~al.}(2013)Napolitano, Capponi, and
  Vanroy]{Napolitano:2013rc}
Napolitano,~S.; Capponi,~S.; Vanroy,~B. Glassy dynamics of soft matter under 1D
  confinement: How irreversible adsorption affects molecular packing, mobility
  gradients and orientational polarization in thin films. \emph{The European
  Physical Journal E} \textbf{2013}, \emph{36}, 61, DOI:
  \doi{10.1140/epje/i2013-13061-8}\relax
\mciteBstWouldAddEndPuncttrue
\mciteSetBstMidEndSepPunct{\mcitedefaultmidpunct}
{\mcitedefaultendpunct}{\mcitedefaultseppunct}\relax
\EndOfBibitem
\bibitem[Napolitano \latin{et~al.}(2017)Napolitano, Glynos, and
  Tito]{Napolitano_2017}
Napolitano,~S.; Glynos,~E.; Tito,~N.~B. Glass transition of polymers in bulk,
  confined geometries, and near interfaces. \emph{Reports on Progress in
  Physics} \textbf{2017}, \emph{80}, 036602, DOI:
  \doi{10.1088/1361-6633/aa5284}\relax
\mciteBstWouldAddEndPuncttrue
\mciteSetBstMidEndSepPunct{\mcitedefaultmidpunct}
{\mcitedefaultendpunct}{\mcitedefaultseppunct}\relax
\EndOfBibitem
\bibitem[Tracht \latin{et~al.}(1998)Tracht, Wilhelm, Heuer, Feng, Schmidt-Rohr,
  and Spiess]{TrachtSpiessDynHetPRL98}
Tracht,~U.; Wilhelm,~M.; Heuer,~A.; Feng,~H.; Schmidt-Rohr,~K.; Spiess,~H.~W.
  Length Scale of Dynamic Heterogeneities at the Glass Transition Determined by
  Multidimensional Nuclear Magnetic Resonance. \emph{Phys. Rev. Lett.}
  \textbf{1998}, \emph{81}, 2727--2730\relax
\mciteBstWouldAddEndPuncttrue
\mciteSetBstMidEndSepPunct{\mcitedefaultmidpunct}
{\mcitedefaultendpunct}{\mcitedefaultseppunct}\relax
\EndOfBibitem
\bibitem[Colmenero \latin{et~al.}(2002)Colmenero, Alvarez, and
  Arbe]{Colmenero_MD_DH_PRE02}
Colmenero,~J.; Alvarez,~F.; Arbe,~A. Self-motion and the \ensuremath{\alpha}
  relaxation in a simulated glass-forming polymer: Crossover from Gaussian to
  non-Gaussian dynamic behavior. \emph{Phys. Rev. E} \textbf{2002}, \emph{65},
  041804, DOI: \doi{10.1103/PhysRevE.65.041804}\relax
\mciteBstWouldAddEndPuncttrue
\mciteSetBstMidEndSepPunct{\mcitedefaultmidpunct}
{\mcitedefaultendpunct}{\mcitedefaultseppunct}\relax
\EndOfBibitem
\bibitem[Weeks \latin{et~al.}(2000)Weeks, Crocker, Levitt, Schofield, and
  Weitz]{WeeksDHColloidScience00}
Weeks,~E.~R.; Crocker,~J.~C.; Levitt,~A.~C.; Schofield,~A.; Weitz,~D.~A.
  Three-Dimensional Direct Imaging of Structural Relaxation Near the Colloidal
  Glass Transition. \emph{Science} \textbf{2000}, \emph{287}, 627--631, DOI:
  \doi{10.1126/science.287.5453.627}\relax
\mciteBstWouldAddEndPuncttrue
\mciteSetBstMidEndSepPunct{\mcitedefaultmidpunct}
{\mcitedefaultendpunct}{\mcitedefaultseppunct}\relax
\EndOfBibitem
\bibitem[Fragiadakis and Roland(2017)Fragiadakis, and
  Roland]{FragiadakisRolandAlfaBetaPRE17}
Fragiadakis,~D.; Roland,~C.~M. Role of structure in the $\ensuremath{\alpha}$
  and $\ensuremath{\beta}$ dynamics of a simple glass-forming liquid.
  \emph{Phys. Rev. E} \textbf{2017}, \emph{95}, 022607, DOI:
  \doi{10.1103/PhysRevE.95.022607}\relax
\mciteBstWouldAddEndPuncttrue
\mciteSetBstMidEndSepPunct{\mcitedefaultmidpunct}
{\mcitedefaultendpunct}{\mcitedefaultseppunct}\relax
\EndOfBibitem
\bibitem[Hansen and McDonald(2006)Hansen, and McDonald]{HansenMcDonaldIIIEd}
Hansen,~J.~P.; McDonald,~I.~R. \emph{Theory of Simple Liquids, 3rd Ed.};
  Academic Press, 2006\relax
\mciteBstWouldAddEndPuncttrue
\mciteSetBstMidEndSepPunct{\mcitedefaultmidpunct}
{\mcitedefaultendpunct}{\mcitedefaultseppunct}\relax
\EndOfBibitem
\bibitem[Zorn(1997)]{ZornNonGaussianPolybutadienePRB97}
Zorn,~R. Deviation from Gaussian behavior in the self-correlation function of
  the proton motion in polybutadiene. \emph{Phys. Rev. B} \textbf{1997},
  \emph{55}, 6249--6259, DOI: \doi{10.1103/PhysRevB.55.6249}\relax
\mciteBstWouldAddEndPuncttrue
\mciteSetBstMidEndSepPunct{\mcitedefaultmidpunct}
{\mcitedefaultendpunct}{\mcitedefaultseppunct}\relax
\EndOfBibitem
\bibitem[Doi and Edwards(1988)Doi, and Edwards]{DoiEdwards}
Doi,~M.; Edwards,~S.~F. \emph{The Theory of Polymer Dynamics}; Clarendon Press,
  1988\relax
\mciteBstWouldAddEndPuncttrue
\mciteSetBstMidEndSepPunct{\mcitedefaultmidpunct}
{\mcitedefaultendpunct}{\mcitedefaultseppunct}\relax
\EndOfBibitem
\bibitem[Bedrov and Smith(2005)Bedrov, and Smith]{BedrovPRE2005}
Bedrov,~D.; Smith,~G.~D. Molecular dynamics simulation study of the $\alpha$
  and $\beta$-relaxation processes in a realistic model polymer. \emph{Physical
  Review E} \textbf{2005}, \emph{71}, 050801\relax
\mciteBstWouldAddEndPuncttrue
\mciteSetBstMidEndSepPunct{\mcitedefaultmidpunct}
{\mcitedefaultendpunct}{\mcitedefaultseppunct}\relax
\EndOfBibitem
\bibitem[Bedrov and Smith(2011)Bedrov, and Smith]{BedrovJNCS2011}
Bedrov,~D.; Smith,~G.~D. Secondary Johari--Goldstein relaxation in linear
  polymer melts represented by a simple bead-necklace model. \emph{Journal of
  Non-Crystalline Solids} \textbf{2011}, \emph{357}, 258--263\relax
\mciteBstWouldAddEndPuncttrue
\mciteSetBstMidEndSepPunct{\mcitedefaultmidpunct}
{\mcitedefaultendpunct}{\mcitedefaultseppunct}\relax
\EndOfBibitem
\bibitem[Fragiadakis and Roland(2017)Fragiadakis, and
  Roland]{FragiadakisRolandMM17}
Fragiadakis,~D.; Roland,~C.~M. Participation in the Johari--Goldstein Process:
  Molecular Liquids versus Polymers. \emph{Macromolecules} \textbf{2017},
  \emph{50}, 4039--4042\relax
\mciteBstWouldAddEndPuncttrue
\mciteSetBstMidEndSepPunct{\mcitedefaultmidpunct}
{\mcitedefaultendpunct}{\mcitedefaultseppunct}\relax
\EndOfBibitem
\bibitem[Richter \latin{et~al.}(1997)Richter, Monkenbusch, Arbe, Colmenero, and
  Farago]{Richter_PhysB1997}
Richter,~D.; Monkenbusch,~M.; Arbe,~A.; Colmenero,~J.; Farago,~B. Dynamic
  structure factors due to relaxation processes in glass-forming polymers.
  \emph{Physica B: Condensed Matter} \textbf{1997}, \emph{241-243},
  1005--1012\relax
\mciteBstWouldAddEndPuncttrue
\mciteSetBstMidEndSepPunct{\mcitedefaultmidpunct}
{\mcitedefaultendpunct}{\mcitedefaultseppunct}\relax
\EndOfBibitem
\bibitem[T\"olle(2001)]{tolle}
T\"olle,~A. Neutron scattering studies of the model glass former
  $ortho$-terphenyl. \emph{Rep. Prog. Phys.} \textbf{2001}, \emph{64},
  1473--1532\relax
\mciteBstWouldAddEndPuncttrue
\mciteSetBstMidEndSepPunct{\mcitedefaultmidpunct}
{\mcitedefaultendpunct}{\mcitedefaultseppunct}\relax
\EndOfBibitem
\bibitem[Fragiadakis and Roland(2014)Fragiadakis, and
  Roland]{Fragiadakis_PRE2014}
Fragiadakis,~D.; Roland,~C.~M. Dynamic correlations and heterogeneity in the
  primary and secondary relaxations of a model molecular liquid. \emph{Physical
  Review E} \textbf{2014}, \emph{89}, 052304--\relax
\mciteBstWouldAddEndPuncttrue
\mciteSetBstMidEndSepPunct{\mcitedefaultmidpunct}
{\mcitedefaultendpunct}{\mcitedefaultseppunct}\relax
\EndOfBibitem
\bibitem[Tripodo \latin{et~al.}(2020)Tripodo, Puosi, Malvaldi, Capaccioli, and
  Leporini]{TripodoBetaPolymers2020}
Tripodo,~A.; Puosi,~F.; Malvaldi,~M.; Capaccioli,~S.; Leporini,~D. Coincident
  Correlation between Vibrational Dynamics and Primary Relaxation of Polymers
  with Strong or Weak Johari-Goldstein Relaxation. \emph{Polymers}
  \textbf{2020}, \emph{12}, 761, DOI: \doi{10.3390/polym12040761}\relax
\mciteBstWouldAddEndPuncttrue
\mciteSetBstMidEndSepPunct{\mcitedefaultmidpunct}
{\mcitedefaultendpunct}{\mcitedefaultseppunct}\relax
\EndOfBibitem
\bibitem[Barbieri \latin{et~al.}(2004)Barbieri, Campani, Capaccioli, and
  Leporini]{CapacciEtAl04}
Barbieri,~A.; Campani,~E.; Capaccioli,~S.; Leporini,~D. Molecular dynamics
  study of the thermal and the density effects on the local and the large-scale
  motion of polymer melts: Scaling properties and dielectric relaxation.
  \emph{J. Chem. Phys.} \textbf{2004}, \emph{120}, 437--453\relax
\mciteBstWouldAddEndPuncttrue
\mciteSetBstMidEndSepPunct{\mcitedefaultmidpunct}
{\mcitedefaultendpunct}{\mcitedefaultseppunct}\relax
\EndOfBibitem
\bibitem[De{ }Michele and Leporini(2001)De{ }Michele, and
  Leporini]{CristianoSE}
De{ }Michele,~C.; Leporini,~D. Viscous flow and jump dynamics in molecular
  supercooled liquids. I. Translations. \emph{Phys. Rev. E} \textbf{2001},
  \emph{63}, 036701\relax
\mciteBstWouldAddEndPuncttrue
\mciteSetBstMidEndSepPunct{\mcitedefaultmidpunct}
{\mcitedefaultendpunct}{\mcitedefaultseppunct}\relax
\EndOfBibitem
\bibitem[Ottochian \latin{et~al.}(2008)Ottochian, De{ }Michele, and
  Leporini]{NGP_Wolynes}
Ottochian,~A.; De{ }Michele,~C.; Leporini,~D. Non-Gaussian effects in the cage
  dynamics of polymers. \emph{Philosophical Magazine} \textbf{2008}, \emph{88},
  4057--4062\relax
\mciteBstWouldAddEndPuncttrue
\mciteSetBstMidEndSepPunct{\mcitedefaultmidpunct}
{\mcitedefaultendpunct}{\mcitedefaultseppunct}\relax
\EndOfBibitem
\bibitem[Cui \latin{et~al.}(2018)Cui, Evenson, Fan, Li, Wang, and
  Zaccone]{ZacconeBetaPRB18}
Cui,~B.; Evenson,~Z.; Fan,~B.; Li,~M.-Z.; Wang,~W.-H.; Zaccone,~A. Possible
  origin of $\ensuremath{\beta}$-relaxation in amorphous metal alloys from
  atomic-mass differences of the constituents. \emph{Phys. Rev. B}
  \textbf{2018}, \emph{98}, 144201, DOI: \doi{10.1103/PhysRevB.98.144201}\relax
\mciteBstWouldAddEndPuncttrue
\mciteSetBstMidEndSepPunct{\mcitedefaultmidpunct}
{\mcitedefaultendpunct}{\mcitedefaultseppunct}\relax
\EndOfBibitem
\bibitem[Cui \latin{et~al.}(2018)Cui, Gebbia, Tamarit, and
  Zaccone]{ZacconeTamaritBetaPRE18}
Cui,~B.; Gebbia,~J.~F.; Tamarit,~J.-L.; Zaccone,~A. Disentangling
  $\ensuremath{\alpha}$ and $\ensuremath{\beta}$ relaxation in orientationally
  disordered crystals with theory and experiments. \emph{Phys. Rev. E}
  \textbf{2018}, \emph{97}, 053001, DOI: \doi{10.1103/PhysRevE.97.053001}\relax
\mciteBstWouldAddEndPuncttrue
\mciteSetBstMidEndSepPunct{\mcitedefaultmidpunct}
{\mcitedefaultendpunct}{\mcitedefaultseppunct}\relax
\EndOfBibitem
\bibitem[Cui \latin{et~al.}(2020)Cui, Gebbia, Romanini,
  Rudi\ifmmode~\acute{c}\else \'{c}\fi{}, Fernandez-Perea, Bermejo, Tamarit,
  and Zaccone]{ZacconeTamaritBetaPRB2020}
Cui,~B.; Gebbia,~J.~F.; Romanini,~M.; Rudi\ifmmode~\acute{c}\else
  \'{c}\fi{},~S.; Fernandez-Perea,~R.; Bermejo,~F.~J.; Tamarit,~J.-L.;
  Zaccone,~A. Secondary relaxation in the terahertz range in 2-adamantanone
  from theory and experiments. \emph{Phys. Rev. B} \textbf{2020}, \emph{101},
  104202, DOI: \doi{10.1103/PhysRevB.101.104202}\relax
\mciteBstWouldAddEndPuncttrue
\mciteSetBstMidEndSepPunct{\mcitedefaultmidpunct}
{\mcitedefaultendpunct}{\mcitedefaultseppunct}\relax
\EndOfBibitem
\bibitem[Zaccone(2020)]{Zaccone_ReviewRelaxVibratGlasses_JPCM2020}
Zaccone,~A. Relaxation and vibrational properties in metal alloys and other
  disordered systems. \emph{Journal of Physics: Condensed Matter}
  \textbf{2020}, \emph{32}, 203001, DOI: \doi{10.1088/1361-648x/ab6e41}\relax
\mciteBstWouldAddEndPuncttrue
\mciteSetBstMidEndSepPunct{\mcitedefaultmidpunct}
{\mcitedefaultendpunct}{\mcitedefaultseppunct}\relax
\EndOfBibitem
\bibitem[Widmer-Cooper and Harrowell(2006)Widmer-Cooper, and
  Harrowell]{Harrowell06}
Widmer-Cooper,~A.; Harrowell,~P. Predicting the Long-Time Dynamic Heterogeneity
  in a Supercooled Liquid on the Basis of Short-Time Heterogeneities.
  \emph{Phys. Rev. Lett.} \textbf{2006}, \emph{96}, 185701\relax
\mciteBstWouldAddEndPuncttrue
\mciteSetBstMidEndSepPunct{\mcitedefaultmidpunct}
{\mcitedefaultendpunct}{\mcitedefaultseppunct}\relax
\EndOfBibitem
\bibitem[Larini \latin{et~al.}(2008)Larini, Ottochian, De{ }Michele, and
  Leporini]{OurNatPhys}
Larini,~L.; Ottochian,~A.; De{ }Michele,~C.; Leporini,~D. Universal scaling
  between structural relaxation and vibrational dynamics in glass-forming
  liquids and polymers. \emph{Nature Physics} \textbf{2008}, \emph{4},
  42--45\relax
\mciteBstWouldAddEndPuncttrue
\mciteSetBstMidEndSepPunct{\mcitedefaultmidpunct}
{\mcitedefaultendpunct}{\mcitedefaultseppunct}\relax
\EndOfBibitem
\bibitem[Puosi \latin{et~al.}(2019)Puosi, Tripodo, and Leporini]{ReviewIJMS}
Puosi,~F.; Tripodo,~A.; Leporini,~D. Fast Vibrational Modes and Slow
  Heterogeneous Dynamics in Polymers and Viscous Liquids. \emph{International
  Journal of Molecular Sciences} \textbf{2019}, \emph{20}, 5708\relax
\mciteBstWouldAddEndPuncttrue
\mciteSetBstMidEndSepPunct{\mcitedefaultmidpunct}
{\mcitedefaultendpunct}{\mcitedefaultseppunct}\relax
\EndOfBibitem
\bibitem[Plimpton(1995)]{PlimptonLAMMPS}
Plimpton,~S. Fast Parallel Algorithms for Short-Range Molecular Dynamics.
  \emph{J. Comput. Phys.} \textbf{1995}, \emph{117}, 1--19\relax
\mciteBstWouldAddEndPuncttrue
\mciteSetBstMidEndSepPunct{\mcitedefaultmidpunct}
{\mcitedefaultendpunct}{\mcitedefaultseppunct}\relax
\EndOfBibitem
\bibitem[Pli()]{PlimptonURL}
http://lammps.sandia.gov\relax
\mciteBstWouldAddEndPuncttrue
\mciteSetBstMidEndSepPunct{\mcitedefaultmidpunct}
{\mcitedefaultendpunct}{\mcitedefaultseppunct}\relax
\EndOfBibitem
\bibitem[Kremer \latin{et~al.}(1988)Kremer, Grest, and
  Carmesin]{KremerGrestCarmesinPRL88}
Kremer,~K.; Grest,~G.~S.; Carmesin,~I. Crossover from Rouse to Reptation
  Dynamics: A Molecular-Dynamics Simulation. \emph{Phys. Rev. Lett.}
  \textbf{1988}, \emph{61}, 566--569\relax
\mciteBstWouldAddEndPuncttrue
\mciteSetBstMidEndSepPunct{\mcitedefaultmidpunct}
{\mcitedefaultendpunct}{\mcitedefaultseppunct}\relax
\EndOfBibitem
\end{mcitethebibliography}

\end{document}